\documentclass[a4paper,11pt]{article}
\usepackage{pos}
\usepackage{isotope}
\usepackage{xspace}
\usepackage{wrapfig}
\usepackage{rviewport}
\usepackage[capitalize]{cleveref}

\newcommand{\eV}{\ensuremath{\mbox{e\kern-0.1em V}}\xspace}
\newcommand{\GeV}{\ensuremath{\mbox{Ge\kern-0.1em V}}\xspace}
\newcommand{\TeV}{\ensuremath{\mbox{Te\kern-0.1em V}}\xspace}
\newcommand{\MeV}{\ensuremath{\mbox{Me\kern-0.1em V}}\xspace}
\newcommand{\GeVc}{\ensuremath{\mbox{Ge\kern-0.1em V}\kern-0.1em/\kern-0.05em c}\xspace}
\newcommand{\GeVcc}{\ensuremath{\mbox{Ge\kern-0.1em V}\kern-0.1em/\kern-0.05em c^2}\xspace}
\newcommand{\AGeV}{\ensuremath{A\,\mbox{Ge\kern-0.1em V}}\xspace}
\newcommand{\AGeVc}{\ensuremath{A\,\mbox{Ge\kern-0.1em V}\kern-0.1em/\kern-0.05em c}\xspace}
\newcommand{\MeVc}{\ensuremath{\mbox{Me\kern-0.1em V}\kern-0.1em/\kern-0.05em c}\xspace}
\newcommand{\MeVcc}{\ensuremath{\mbox{Me\kern-0.1em V}\kern-0.1em/\kern-0.05em c^2}\xspace}

\def \dedx{d$E$/d$x$\xspace}

\newcommand{\NASixtyOne}{NA61/\-SHINE\xspace}  %this adds a '-' sign in case of line break
\newcommand{\CernVM}{\textsc{Cern\-\kern-0.05emVM}\xspace}

\begin{document}

\title{Results from a Pilot Study on the Measurement of Nuclear Fragmentation with \NASixtyOne at the CERN SPS:\\ $^\text{11}$C Production in C+p Interactions at 13.5\,\textit{A}\,GeV/\textit{c}}
\ShortTitle{Fragmentation Cross Section Measurement with \NASixtyOne}
\author*{Neeraj Amin}
\affiliation{Institute for Astroparticle Physics, Karlsruhe Institute of Technology~\includegraphics[height=1.55ex]{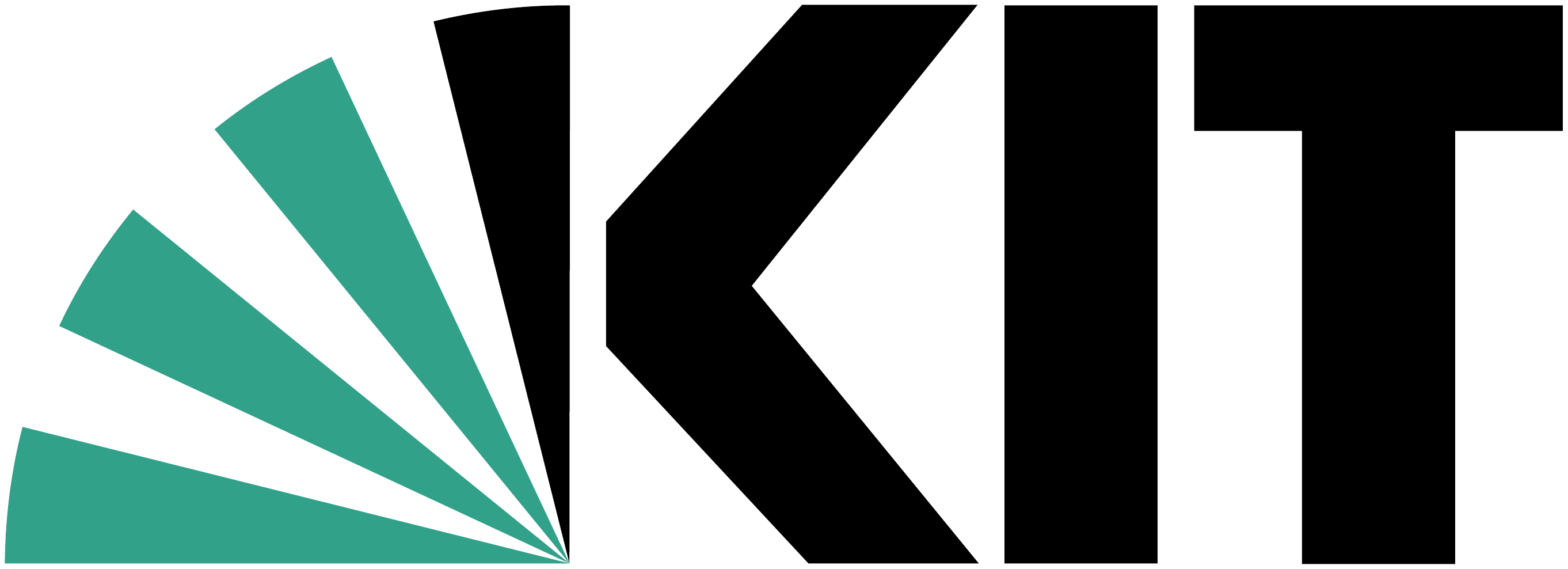}, Germany}
\forColl{\NASixtyOne}
\emailAdd{neeraj.amin@kit.edu}

\abstract{
We report the analysis of data taken during a pilot run in 2018 to study the feasibility of nuclear fragmentation measurements with the \NASixtyOne experiment at the CERN SPS.
These nuclear reactions are important for the interpretation of secondary cosmic-ray nuclei production (Li, Be, and B) in the Galaxy.

The pilot data were taken with \isotope[12]{C} projectiles at a beam momentum of 13.5\,\AGeVc and two fixed targets, polyethylene (C$_2$H$_4$) and graphite.
The specific focus here is the measurement of total Boron (\isotope[10]{B} and \isotope[11]{B}) production cross section in C+p interactions at 13.5\,\AGeVc.
The cosmic-ray nucleus \isotope[11]{C} is termed a `Ghost nucleus' on account of its short lifetime compared to the usual cosmic-ray diffusion time in the Galaxy and it ultimately decays to Boron as, $\isotope[11]{C} \to \isotope[11]{B} + \beta^+$.
Therefore, precise knowledge of the production cross section of \isotope[11]{C} is very relevant for the understanding of Boron production in the Galaxy.
We present a preliminary measurement of the fragmentation cross section of $\isotope{C} + \text{p} \to \isotope[11]{C}$, which, together with our previously reported B-production cross section, provides a new constraint on boron production in the Galaxy in the high-energy range relevant for modern space based cosmic-ray experiments like AMS-02.
}

\FullConference{37$^\text{th}$ International Cosmic Ray Conference (ICRC 2021)\\
		July 12th -- 23rd, 2021\\
		Online -- Berlin, Germany}

%The \NASixtyOne facility is a multi-purpose experiment
%  located at the H2 beam line of the CERN SPS North Area, with the aim
%  of studying the properties of hadron production in nuclear
%  collisions with fixed targets. Important goals are to measure
%  hadron-nucleus interactions to improve cosmic-ray shower modeling
%  and also to study light, secondary cosmic-ray nuclei production (Li,
%  Be \& B) in the Galaxy.
%\tableofcontents%

%\begin{document}
\maketitle

\section{Introduction}
Cosmic-ray (CR) propagation in the Galaxy can be constrained by modeling the secondary-to-primary flux ratios of cosmic rays at Earth, for instance the Boron-to-Carbon flux ratio.

Space-based CR detectors have recently reported improved measurements of the ratio of secondary to primary Galactic cosmic rays (GCRs) with a precision of ${<}5\%$ at energies of $\gtrsim$10~\GeV~\cite{Adriani:2014xoa,AMS:2016brs,dampeICRC,caletICRC}.
However, especially at these high energies, our insufficient knowledge of the fragmentation cross sections limits the predictive power of CR propagation calculations leading to uncertainties of up to 20\%~\cite{geno}.
Since the current cosmic-ray measurements cover energies up to several hundreds of \GeV, it is therefore important to improve the accuracy of laboratory measurement of the fragmentation cross section above 10\,\GeVc per nucleon, to reduce uncertainties in GCR propagation models.

An important reaction for the diagnostics of CR propagation is the fragmentation of primary CR Carbon (\isotope[12]{C}) to Boron (\isotope[11]{B} and \isotope[10]{B}) when it interacts with the interstellar medium (ISM), which mainly consists of protons.
Results on the direct Boron production with \NASixtyOne have been reported in Ref.~\cite{munger1}.
Here we focus on the additional contribution to B-production in the Galaxy originating from the unstable \isotope[11]{C} isotopes that decay to \isotope[11]{B} via $\isotope[11]{C}^* \to \isotope[11]{B} + \beta^+$.
\isotope[11]{C} has a short lifetime compared to cosmic ray propagation timescales in the Galaxy.
Hence, $\isotope[12]{C} + \text{p} \to \isotope[11]{C}$ production cross section is important for deriving propagation characteristics of CRs.

\begin{figure}[t]
\centering
\includegraphics[height=0.5\linewidth]{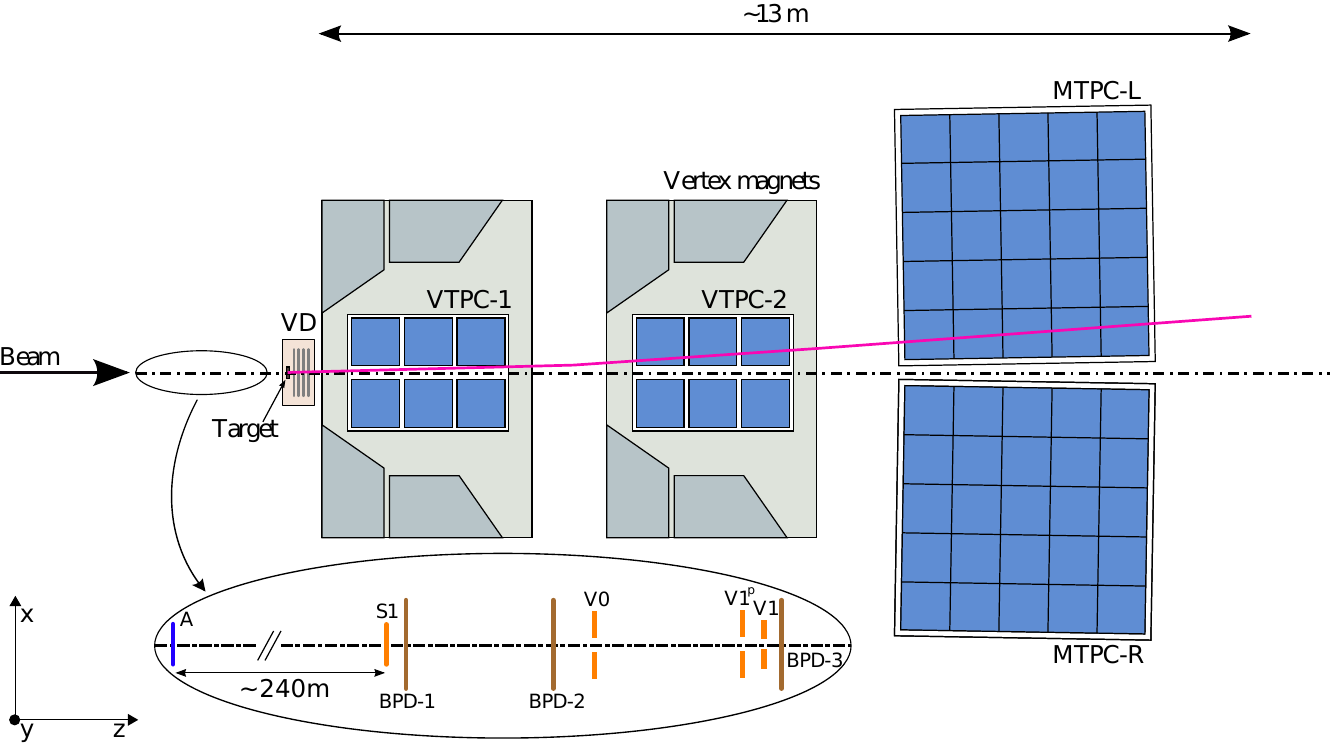}
\caption{\NASixtyOne detector setup for the 2018 pilot run on nuclear fragmentation.}
\label{fig:1}
\end{figure}

\section{\NASixtyOne}

The \NASixtyOne (SPS Heavy Ion and Neutrino Experiment) facility is a multi-purpose experiment located at the H2 beam line of the CERN Super Proton Synchrotron (SPS).
The main aim of the experiment is to study properties of hadron production in nuclear interactions at SPS energies with fixed targets.
The facility provides a unique opportunity for the measurement of the production of light secondary nuclei, like Li, Be, and B, by studying fragmentation of light nuclei, like C, N, and O, relevant to GCR propagation~\cite{na61CR}.
For this purpose a primary $\isotope{Pb}^{82+}$ ion beam from the SPS can be collided with the primary T2 Beryllium target of the H2 beam line.
Fragmentation of the Pb nucleus produces secondary nuclei which are then rigidity-selected and transported to the \NASixtyOne facility that is shown in \cref{fig:1}.
Three Beam Position Detectors (BPDs) along with veto scintillator detectors (V0, V1, and V1$^\text{P}$) are placed upstream of the target to measure the beam direction and define the beam trigger, respectively.
The main sub-detectors used for the measurements are time projection chambers (TPCs).
The two Vertex TPCs (VTPCs) are placed inside two superconducting magnets with a total bending power of 9\,Tm followed by two Main TPCs (MTPCs) after the magnets.
For further details on the experimental setup see Ref.~\cite{na61}.

\section{Measurement with \NASixtyOne}

\subsection{2018 Pilot Run}

\begin{figure}[t]
\centering
\def\h{0.25}
\includegraphics[height=\h\linewidth]{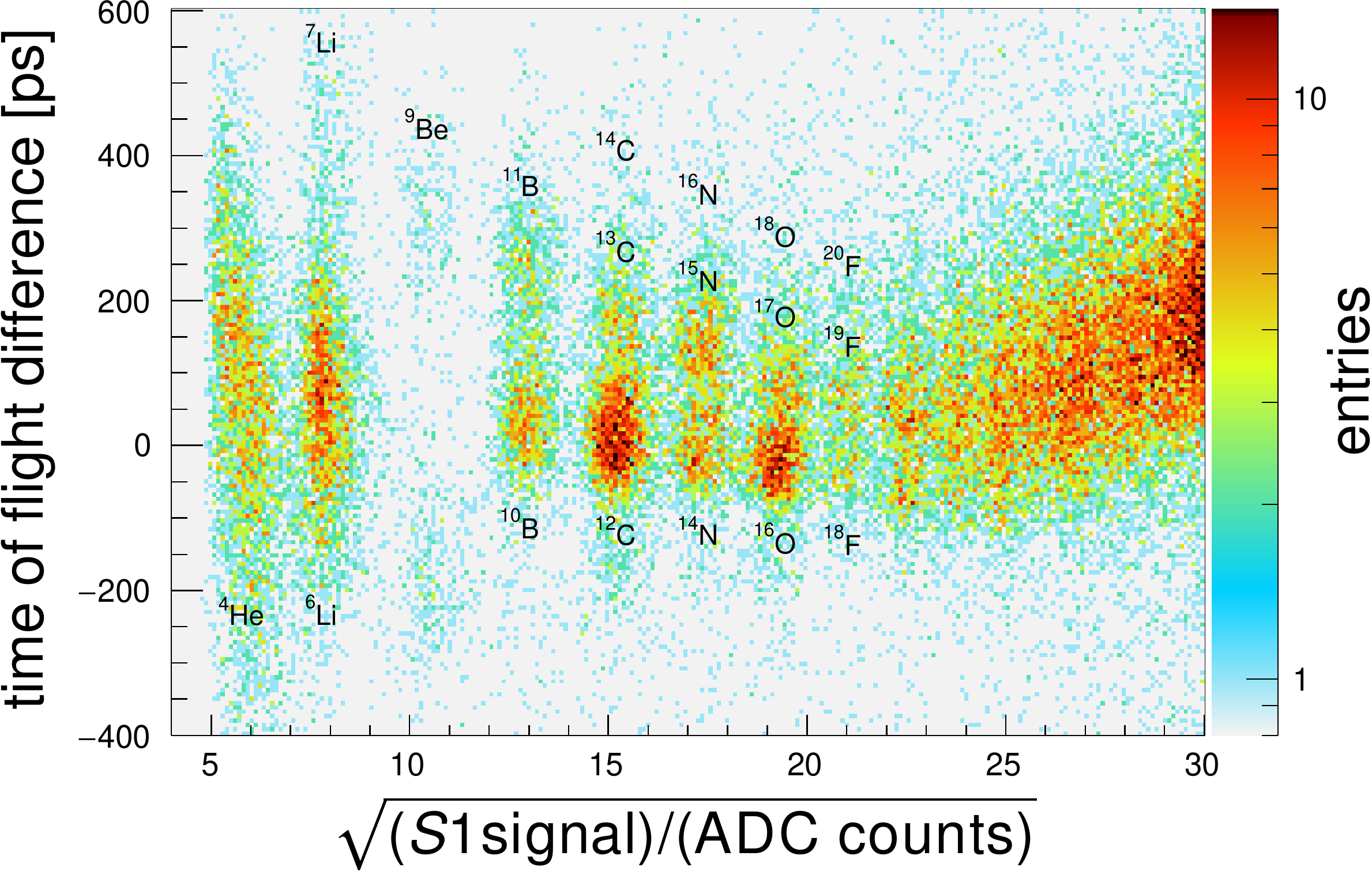}\hfill
\includegraphics[clip,rviewport=-0.05 -0.48 1.05 1,width=0.23\linewidth]{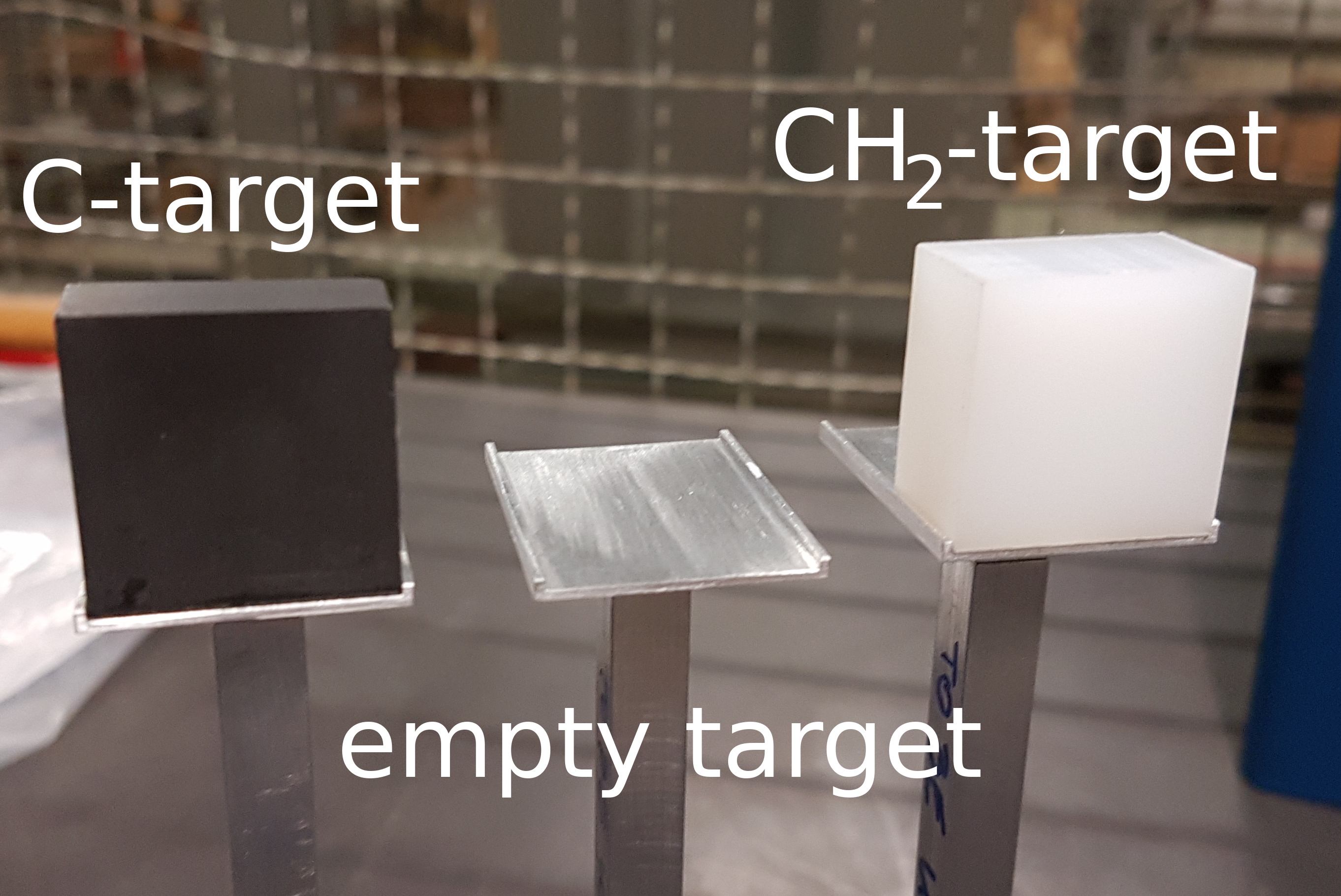}\hfill
\includegraphics[height=\h\linewidth]{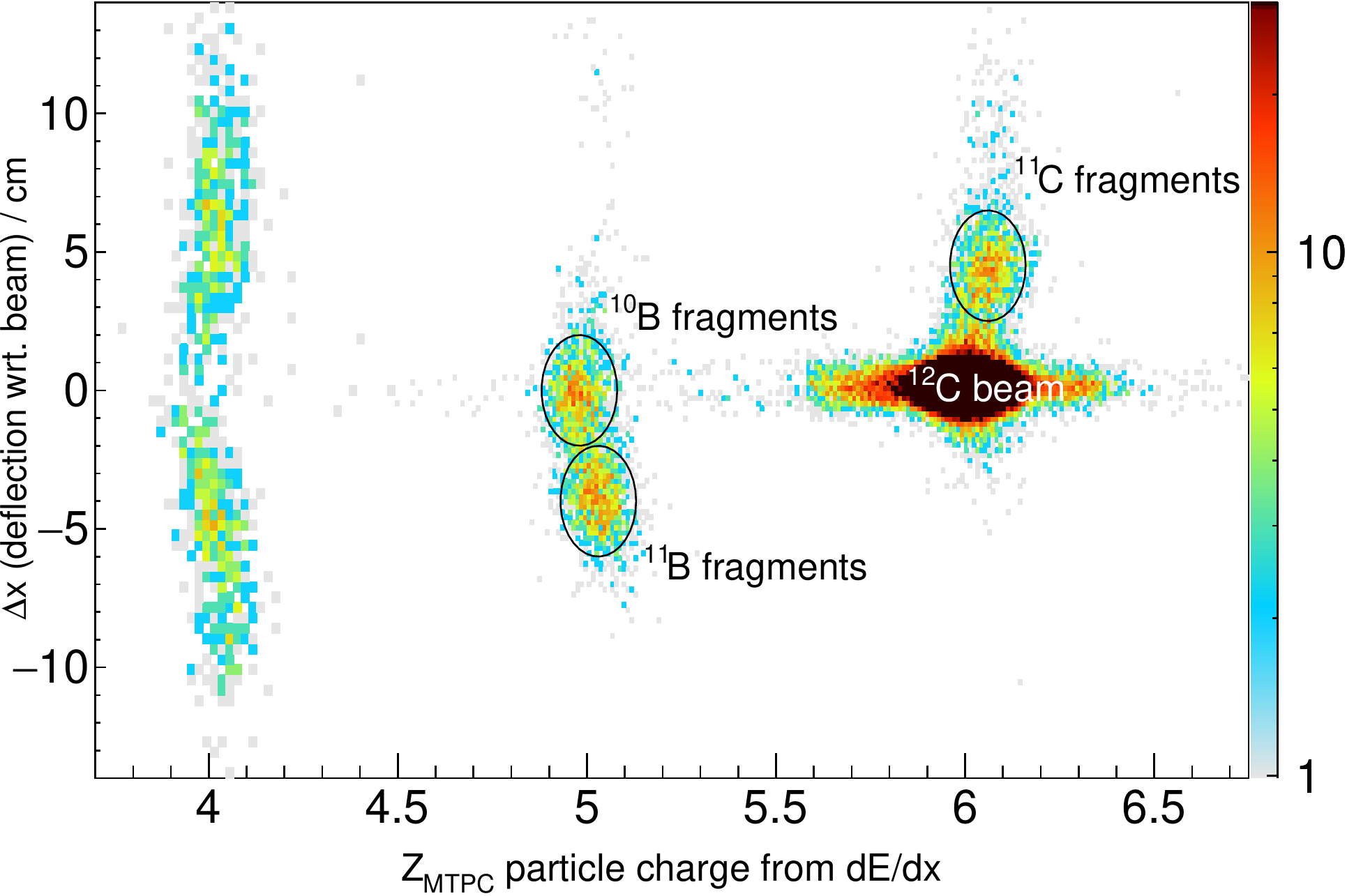}
\caption{\emph{Left:} secondary ion beam composition delivered to NA61;
%  shows nuclei separated
%  by time-of-flight measured between the A and S1 scintillators, on
%  the vertical axis, whereas the horizontal axis shows the charge
% equivalent energy deposit in S1; middle: the target holder with C,
\emph{Middle:} C, OUT, and PE targets;
\emph{Right:} fragments as seen in the MTPC.}
\label{fig:2}
\end{figure}

A pilot run dedicated to measuring nuclear fragmentation reactions was preformed in 2018, recording ${\sim}10^6$ events in the 3-day active data taking period~\cite{pilot}.
It was aimed to study the feasibility of measuring light nuclei production like Boron in \isotope[12]{C}+p interactions.
The secondary ion beam for this run was produced by impinging the primary Pb ion beam on a 16\,cm long Beryllium plate (T2 target) to maximize the \isotope[12]{C} yield.
The collimators and spectrometers on the H2 beam line were tuned to select nuclei with $A/Z=2$ at beam momentum of 13.5\,\GeVc per nucleon.
Beam particles are identified based on the time-of-flight measured between the two scintillators called A and S1, placed ${\sim}240$\,m upstream of the target and the energy deposited in the S1 scintillator.
\cref{fig:2} (left panel) illustrates the beam composition measured with these two quantities.
The beam trigger was set on a charge of 6, and \isotope[12]{C} isotopes were selected offline during data analysis using the time-of-flight measurement.

%The upstream trigger logic was set to (S1
%\ma{\wedge \overline{V0} \wedge \overline{V1} \wedge
%  \overline{V1^P}})~\cite{fsutter}.

For the rest of the article, we refer to \isotope[12]{C} as C in all reactions for convenience.
C+p fragmentation reactions were derived from the measurements with two targets: a 1.5\,cm thick Polyethylene (C$_2$H$_4$, PE) target and a 1.0\,cm thick graphite target (C) to subtract all C+C interactions inside the PE target.
Finally, to correct for interactions of the beam particle in the detectors and other experimental structures, events without any target (henceforth called OUT) were also recorded.
The 3 target settings are shown in the middle panel of \cref{fig:2}.

\subsection{Mass Distribution of Fragments in the MTPC}
\label{frag}

The fragments produced by the beam-target interaction pass through a series of detectors downstream of the target as shown in \cref{fig:1}.
The produced isotopes are identified in the MTPC as shown in the right panel of \cref{fig:2}.
The charge $Z$ of the fragment, shown on the x-axis, is derived from the energy deposit in the chamber gas, which is proportional to $Z^2$.
The deflection of the fragments $\Delta x$ relative to the nominal position of the \isotope[12]{C} beam extrapolated to the chamber after passing through the
magnets depends on the rigidity of the fragment and is hence a measure of $A/Z$, where $A$ denotes the mass number (cf.~\cref{fit}).

To measure production of \isotope[11]{C} from C+p reaction, we select Carbon tracks in the MTPC by placing cuts on $Z_\text{MTPC}$ as $30 \leq Z^2_\text{MTPC} \leq 44.5$.
We perform a likelihood fit to the distribution of Carbon isotopes as a function of $\Delta x$, with an appropriate model.
The $\Delta x$-distributions are shown in \cref{fig:3}.
The three target settings are fitted independently to retrieve individual isotopic yields from each of the data set.
We adopt a flat-top Gaussian with symmetric exponential tails to model the detector response of the beam profile.
Moreover, the fragmented nuclei posses a non-zero Fermi momentum in its rest frame.
The quantum mechanical phenomena induces a sub-GeV scale momentum as a result of the fragmentation, which consequently, when boosted to the laboratory frame and propagated through the magnetic field, produces a spatial distribution of the fragment nuclei as measured in the MTPC.
We modeled this distribution with a Gaussian and convolved it with the detector model for our fits.
The outcome of the fits is discussed in the next section.

\section{Results from the Pilot Run }
\label{res}

\subsection{Fit Results}
\label{fit}

\begin{figure}[t]
\centering
\def\h{0.377}
\includegraphics[height=\h\textwidth]{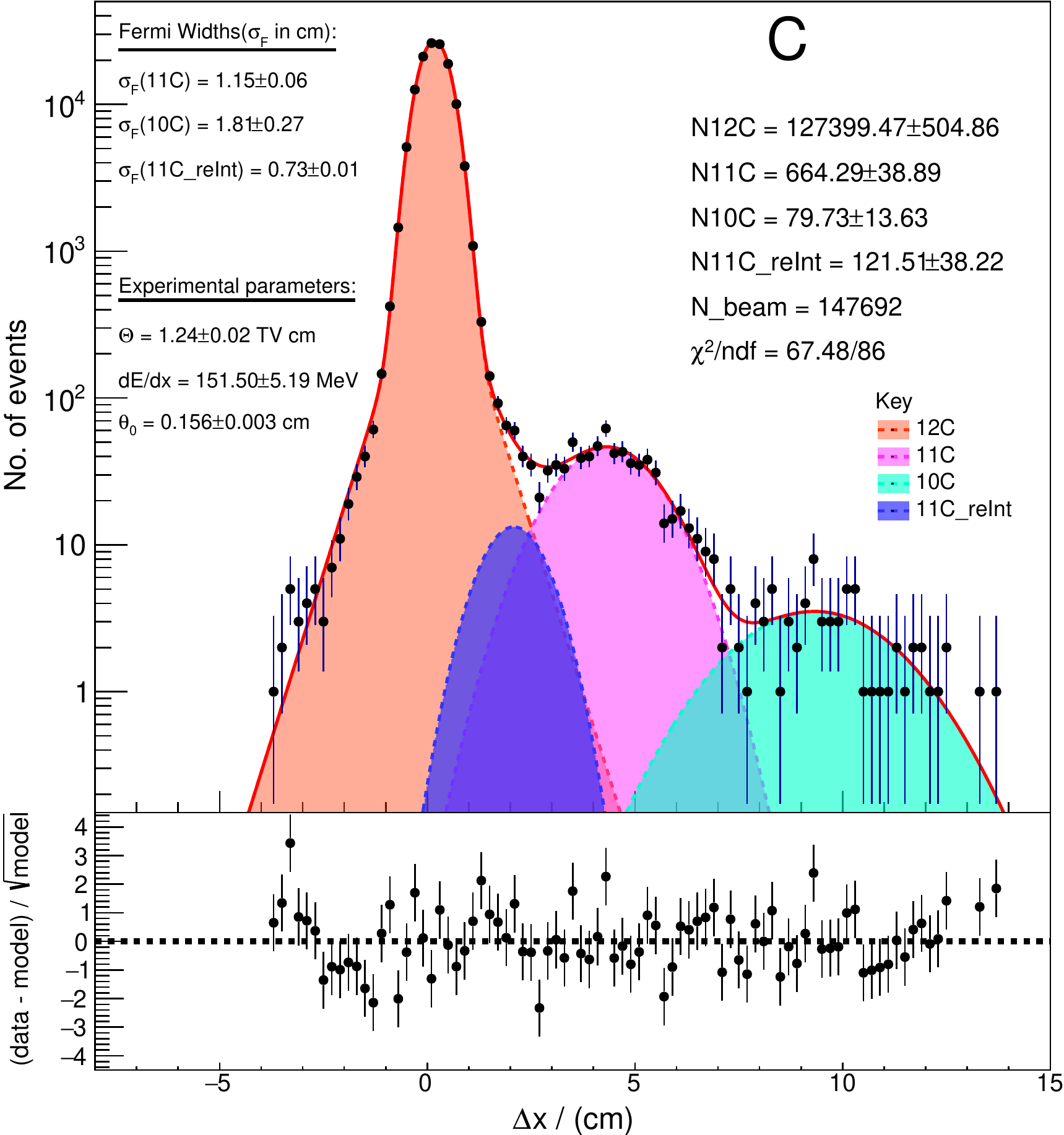}\hfill
\includegraphics[clip,rviewport=0.088 0 1 1,height=\h\textwidth]{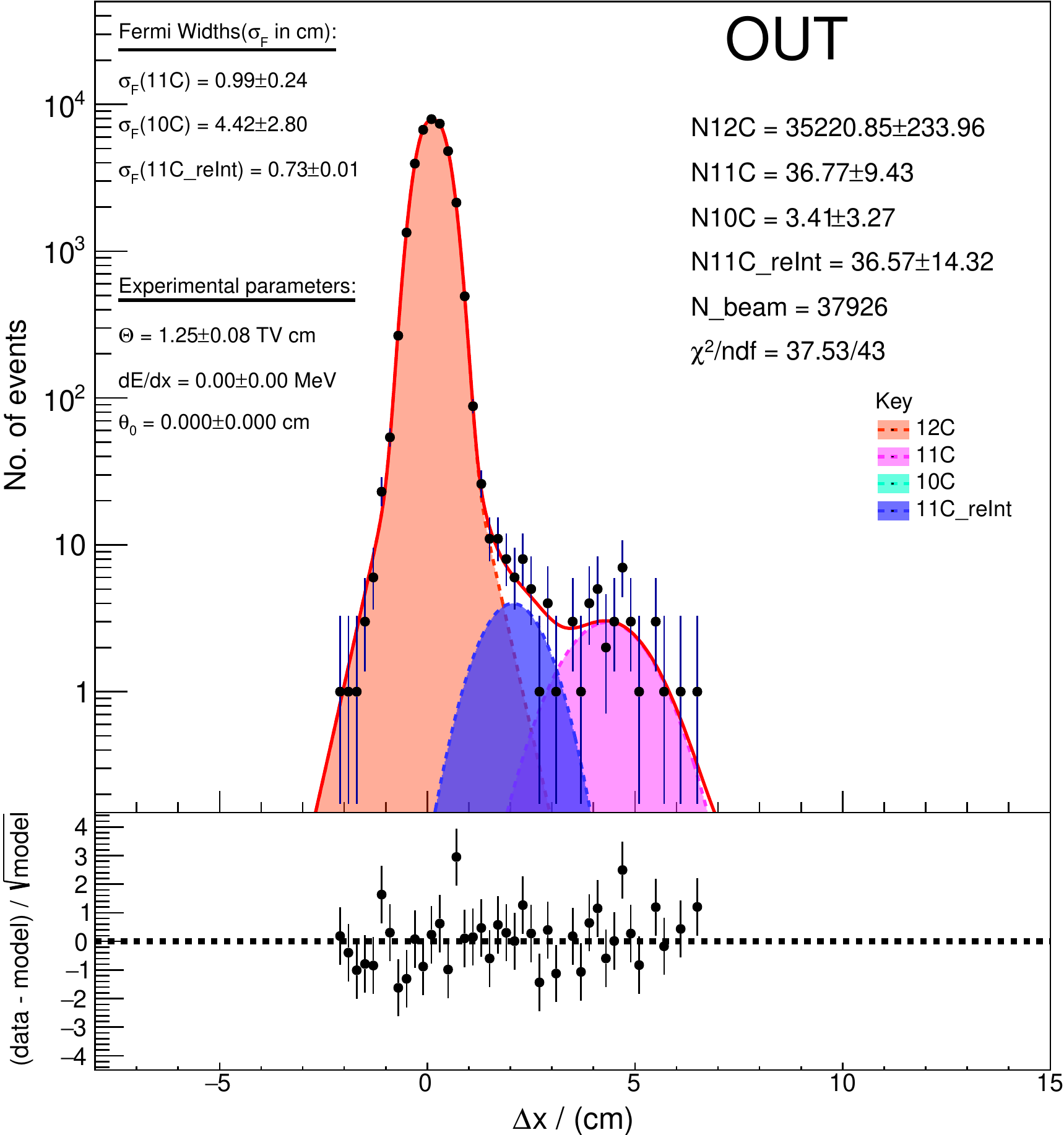}\hfill
\includegraphics[clip,rviewport=0.09 0 1 1,height=\h\textwidth]{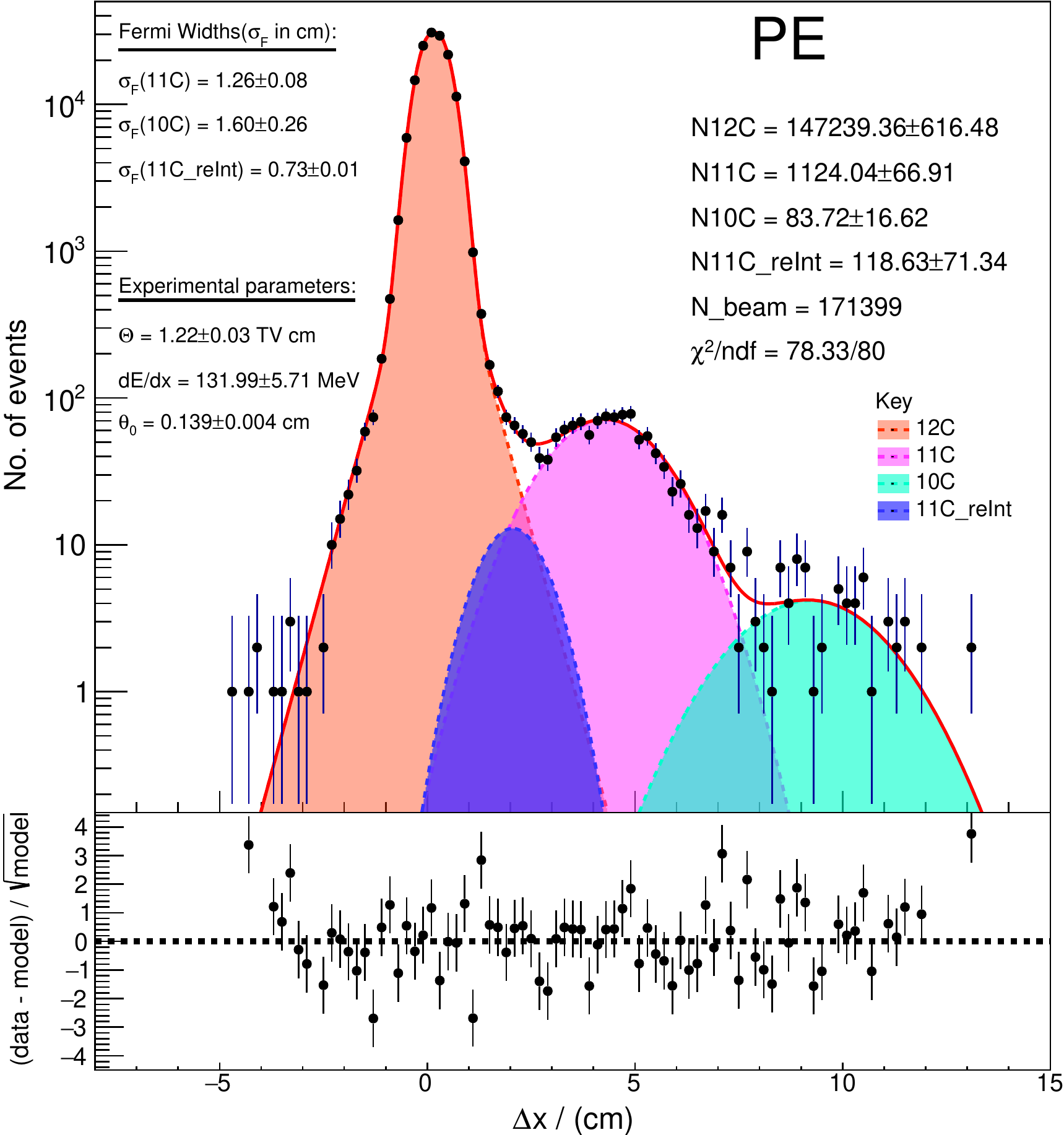}
\caption{Fits of the $\Delta x$-distribution of Carbon isotopes in the MTPC for the three target settings, C (left), empty target (middle), and PE (right).
The yields $N_i$ of the different isotopes are indicated in the top right of each panel, detector-related fit parameters are shown on the top left of each panel (see text for further details).}
\label{fig:3}
\end{figure}

The results of the fits of the $\Delta x$-distributions for the 3 data sets are shown in \cref{fig:3}.
The non-interacting \isotope[12]{C} beam is the central orange peak at $\Delta x \approx 0$, whereas the fragments \isotope[11]{C} and \isotope[10]{C}, deflected to positive $\Delta x$, are shown in magenta and aquamarine, respectively.
In addition to these three peaks, we also included fragments produced in the material between the VTPCs in our fits (represented by the blue peak $\isotope[11]{C}_\text{reInt}$ in \cref{fig:3}).
Parameters corresponding to the isotope widths due to the nuclear Fermi motion of the fragments and those related to experimental effects are indicated on the left side of each plot.

The Fermi momentum derived from the isotope widths, $\sigma_\text{F}$, evaluates to be within 10\% of the known canonical value of ${\sim}250$\,\MeVc.
The deflection of the fragments in the magnetic field is parameterized by $\Theta$, measured in units of TV\,cm.
Since the rigidity of every isotope fragment can be calculated as $R = p_\text{beam} \, A / Z$, their relative positions $\Delta x$ in the MTPC are given by
\begin{equation}
\Delta x = x_\text{iso} - x_\text{beam} = \Theta \, \left(R_\text{iso}^{-1} - R_\text{beam}^{-1}\right),
\end{equation}
where $R_\text{beam} = 27$\,GV for the \isotope[12]{C} beam.
The parameter \dedx quantifies the energy loss of the beam inside the target whereas $\theta_0$ parameterizes the angular broadening due to multiple scattering in the target.
Both the parameters are fixed to 0 in the no-target (OUT) case.

The main result of the fit is the number of isotope fragments reaching the MTPC.
These isotopic yields are shown on the right-side column of the plots in \cref{fig:3}.

\subsection{Calculation of Interaction Probabilities and Cross Sections}

To derive an expression for the interaction cross section in the target it is useful to express the total interaction probability as a function of interaction probabilities in the target itself, and the region up- and downstream of the target~\cite{claudia}.
Then the total measured probability for an interaction $\text{C}\to\text{A}$ of the beam particle can be written as a sum of the interaction probabilities in these 3 regions as follows,
\begin{equation}
P^\text{tot,T}_{\text{C}\to\text{A}} =
  P^\text{up,T}_{\text{C}\to\text{A}} +
  P^\text{intr,T}_{\text{C}\to\text{A}} \left(1-P^\text{up,T}_{\text{C}\to\text{A}}\right) +
  P^\text{down,T}_{\text{C}\to\text{A}} \left(1-P^\text{up,T}_{\text{C}\to\text{A}}\right) \left(1-P^\text{intr,T}_{\text{C}\to\text{A}}\right).
\label{eq:1}
\end{equation}
For the empty target holder case, where $P^\text{intr,T}_{\text{C}\to\text{A}} = 0$, we have,
\begin{equation}
P^\text{tot,OUT}_{\text{C}\to\text{A}} =
  P^\text{up,T}_{\text{C}\to\text{A}} +
  P^\text{down,T}_{\text{C}\to\text{A}}\left(1-P^\text{up,T}_{\text{C}\to\text{A}}\right).
\label{eq:2}
\end{equation}
Here the term interaction (denoted by $\text{C}\to\text{A}$) refers to an inelastic collision wherein the incident nucleus loses more than one nucleon.
The quantity of interest for our measurements is the probability of the beam particle interacting with the target, $P^\text{intr,T}_{\text{C}\to\text{A}}$.
In the above equations, $\text{A}=\{\text{X}, \isotope[11]{C}\}$, where X denotes any nuclei except \isotope[12]{C}, and $\text{T}\equiv\text{Target}=\{\text{PE}, \text{C}\}$.
The total measured interaction probabilities are observables and can be determined directly in terms of the isotope yields from the fit (\cref{fig:3} right column).
For the Carbon inelastic or mass-changing ($\text{C}\to\text{X}$) and \isotope[11]{C} production ($\text{C}\to\isotope[11]{C}$) probabilities, we have,
\begin{equation}
P^\text{tot,T/OUT}_{\text{C}\to\text{X}} = 1 - \frac{N^\text{T/OUT}_{\isotope[12]{C}}}{N^\text{T/OUT}_\text{beam}}
\qquad\text{and}\qquad
P^\text{tot,T/OUT}_{\text{C}\to\isotope[11]{C}} = \frac{N^\text{T/OUT}_{\isotope[11]{C}}}{N^\text{T/OUT}_\text{beam}},
\end{equation}
where, $N^\text{T/OUT}_{\isotope[12]{C}}$ and $N^\text{T/OUT}_{\isotope[11]{C}}$ are retrieved from the fit and $N^\text{T/OUT}_\text{beam}$ is total number of beam particles corresponding to the target setting T/OUT(i.e. target-in or target-out).

Furthermore, for the $\text{C}\to\text{X}$ reaction, \cref{eq:1,eq:2} can be solved simultaneously to obtain an expression for the probability of interaction of the beam particle directly with the target as,
\begin{equation}
P^\text{intr,T}_{\text{C}\to\text{A}} =
  \frac{P^\text{tot,T}_{\text{C}\to\text{A}} - P^\text{tot,OUT}_{\text{C}\to\text{A}}}
       {1 - P^\text{tot,OUT}_{\text{C}\to\text{A}}}.
\label{eq:3}
\end{equation}

The total interaction probability for the $\text{C}\to\isotope[11]{C}$ fragmentation can also be expressed in the form of \cref{eq:1,eq:2}.
However, the system of equations, in this case, is underdetermined.
Thus, an equivalent version of \cref{eq:3} for $\text{C}\to\isotope[11]{C}$ case cannot be determined analytically.
Therefore, we implement a numerical approach, wherein appropriate constraints are applied to the unknowns to calculate the interaction probability of the C-beam with the target, fragmenting to \isotope[11]{C}~\cite{fsutter}.

Finally, the true interaction probability of the projectile with the target can be written in terms of the target thickness $d_\text{T}$ and interaction length, $\lambda$, of the beam particle in the target as~\cite{na61cp09}
\begin{equation}
P^\text{intr,T}_{\text{C}\to\text{A}} =  1 - \exp(-d_\text{T}/\lambda),
\label{eq:4}
\end{equation}
where $\lambda=1/(n_\text{T} \, \sigma^\text{T}_{\text{C}\to\text{A}})$.
The target volume density $n_\text{T} = \rho_\text{T} \, N_\text{A} / M_\text{T}$ is expressed in terms of the density $\rho_\text{T}$, Molar mass $M_\text{T}$, and Avogadro's constant $N_\text{A}$ whereas $\sigma^\text{T}_{\text{C}\to\text{A}}$ is the required cross section.
Making these substitutions and re-arranging \cref{eq:4}, we obtain an expression for the cross section as,
\begin{equation}
\sigma_{\text{C+T}\to\text{A}} =
\sigma^\text{T}_{\text{C}\to\text{A}} =
\frac{-\ln(1 - P^\text{intr,T}_{\text{C}\to\text{A}})}
     {n_\text{T} \, d_\text{T}}.
\label{eq:5}
\end{equation}

Using \cref{eq:5} to calculate interaction cross section for the two targets (PE, C), we can finally compute the cross section for the reaction $\text{C} + \text{p}\to\text{A}$ as
\begin{equation}
\sigma_{\text{C+p}\to\text{A}} =
  \tfrac{1}{2}(\sigma_{\text{C+PE}\to\text{A}} - \sigma_{\text{C+C}\to\text{A}}).
\label{eq:6}
\end{equation}

The factor 2 in the denominator denotes the 2:1 Hydrogen-to-Carbon ratio in Polyethylene.
The measured cross sections are further corrected for upstream beam particle selection and downstream Carbon isotope selection efficiency, re-interaction of the \isotope[11]{C} fragment inside the target and in the detectors as well.
The systematic uncertainties arising due to these corrections and numerical calculation methods are conservatively estimated to be at ${\sim}10\%$ of the statistical error of our measurement.

From this analysis a preliminary cross section for the Carbon mass-changing cross section is computed to be,
\begin{equation}
\sigma_{\text{C+p}\to\text{X}} = 255.3 \pm 10.9\,\text{mb}
\end{equation}
and for the \isotope[11]{C} production cross section we get,
\begin{equation}
\sigma_{\text{C+p}\to\isotope[11]{C}} = 33.4 \pm 3.1\,\text{mb},
\end{equation}
where the uncertainties denote the total uncertainties dominated by statistical uncertainties for this pilot run data.

\begin{figure}[t]
\centering
\def\h{0.36}
\includegraphics[height=\h\linewidth]{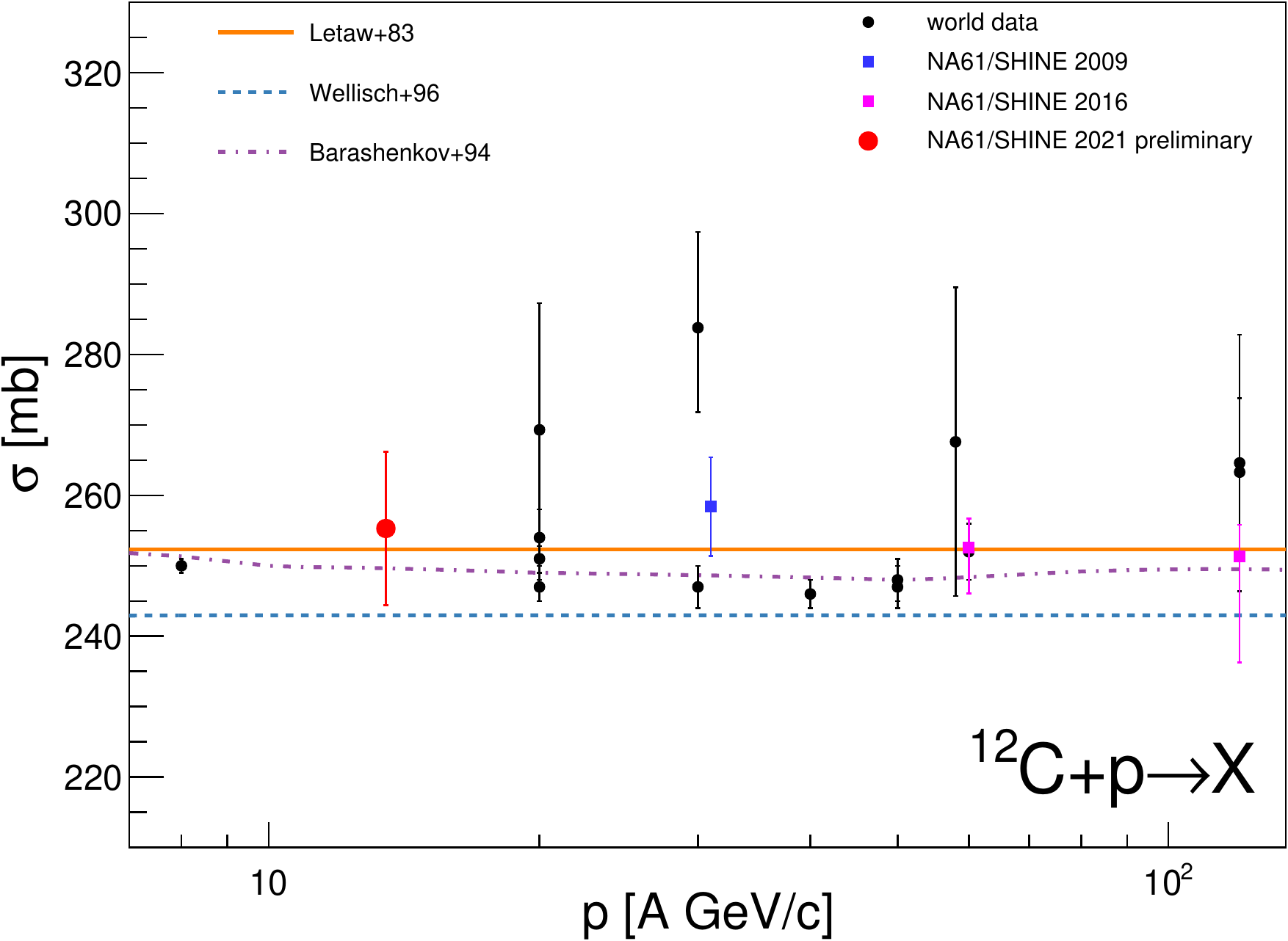}\hfill
\includegraphics[height=\h\linewidth]{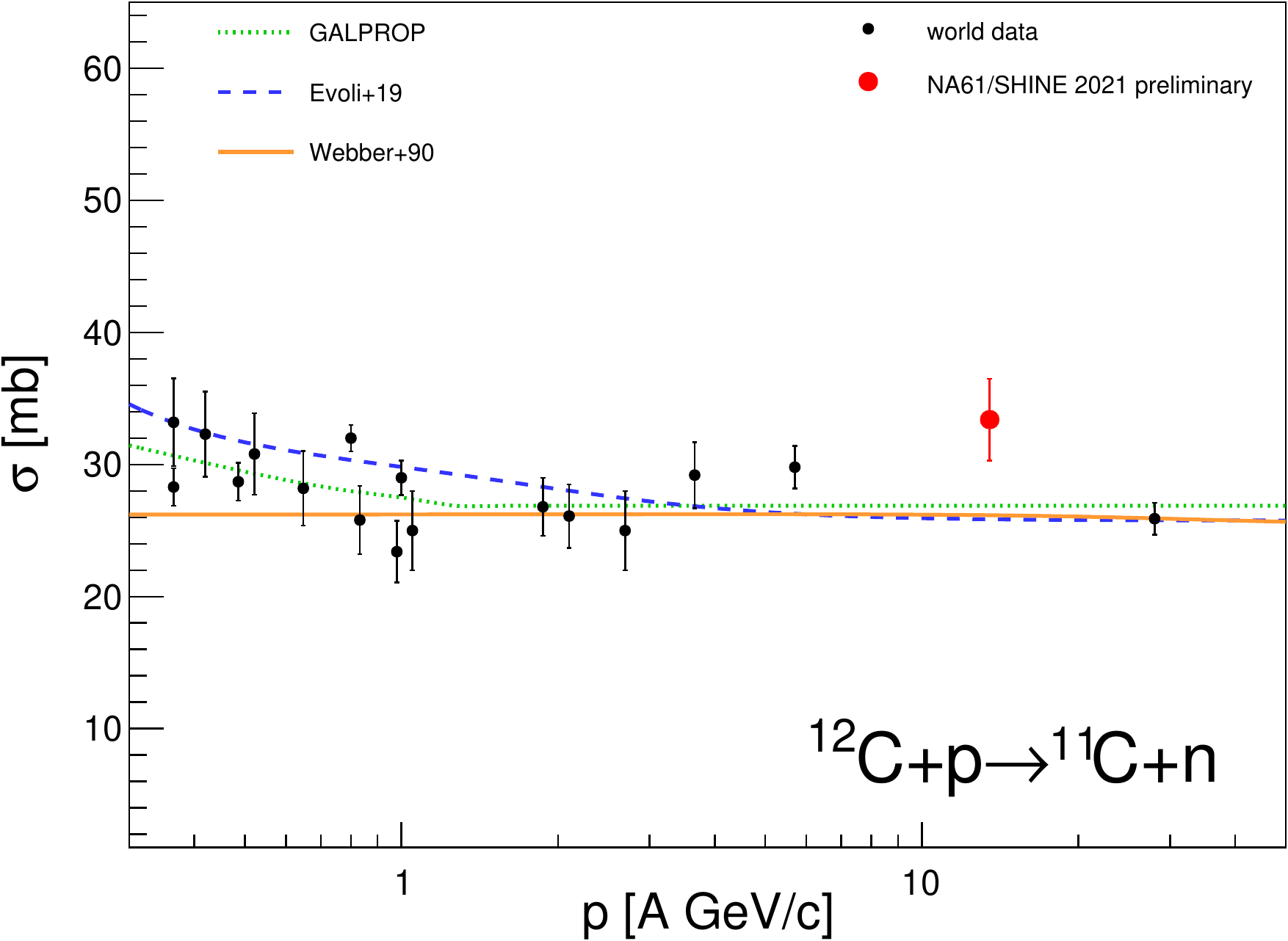}
\caption{\emph{Left:} Preliminary measurement of the $\text{C}+\text{p}\to\text{X}$ reaction compared to previous data \cite{na61cp09, na61cp16, emphatic} and parameterizations \cite{letaw,wellisch,bara}.
\emph{Right:} $\text{C}+\text{p}\to\isotope[11]{C}$ production cross section result compared with  Evoli+19~\cite{evoli}, Webber+90~\cite{wwebber} and GALPROP~\cite{galprop} parameterizations.
%The latter denotes fit results from Ref.~\cite{evoli} based on formula given in Ref.~\cite{reinert}.
Previous data are from Ref.~\cite{evoli} (and references therein).}
\label{fig:4}
\end{figure}

Our measurement results compared with previous data compiled by Refs.~\cite{na61cp09, na61cp16, emphatic, evoli} are shown in \cref{fig:4}.
It is interesting to note that, for the $\text{C}+\text{p}\to\isotope[11]{C}$ reaction, only one data point is available at beam momentum of 28\,\AGeVc~\cite{jbcummings} in the asymptotic region beyond 10\,\AGeVc.
This measurement was performed by irradiating a plastic scintillator target with a proton beam and monitoring the 20.4\,min \isotope[11]{C} activity by the internal scintillation counting method.

\section{Summary and Outlook}
\label{sum}

In this work, we presented a preliminary measurement of the Carbon mass-changing and the \isotope[11]{C} production cross-sections with C+p interaction at 13.5\,\AGeVc from the \NASixtyOne 2018 pilot run for nuclear fragmentation data.
As is evident from \cref{fig:4}, our results are in good overall agreement with previously reported data.
The good resolution on mass and charge, as is apparent from \cref{fig:2}~(right) demonstrates the feasibility of measuring light nuclei fragmentation cross sections relevant to cosmic ray propagation in the galaxy at SPS energies with the \NASixtyOne facility.
The collaboration is in the process of upgrading the detector including an improvement by a factor of 10 of the readout capabilities of data acquisition system.
This will enable us to measure fragmentation of various light nuclei, like C, N, and O, with improved efficiency and high statistics during a data taking dedicated to nuclear fragmentation planned in 2022.

\section*{Acknowledgments}

We would like to thank the organizers for the opportunity to present our results at this conference, ICRC21 and Carmelo Evoli for providing the data tables and fit shown in \cref{fig:4} (right panel).
The fragmentation measurements are supported by the NCN-DFG Beethoven CLASSIC 3 funding of the Deutsche Forschungsgemeinschaft (DFG -- German Research Foundation) project number 426579465 and the Polish Ministry of Science and Higher Education grant 2018/31/G/ST2/03910.

\clearpage

\section*{The \NASixtyOne Collaboration}
\small

\begin{wrapfigure}[6]{l}{0.13\linewidth}
\vspace{-2mm}
\includegraphics[width=0.98\linewidth]{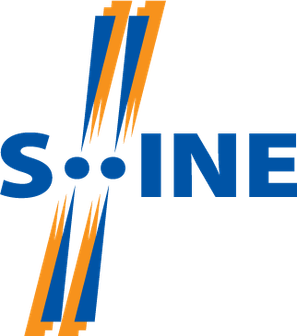}
\end{wrapfigure}
\begin{sloppypar}
% based on XML DB with time Wed Jul 14 09:19:03 2021
% Authors in alphabetical order.
\noindent
A.~Acharya$^{\,10}$,
H.~Adhikary$^{\,10}$,
K.K.~Allison$^{\,26}$,
N.~Amin$^{\,5}$,
E.V.~Andronov$^{\,22}$,
T.~Anti\'ci\'c$^{\,3}$,
V.~Babkin$^{\,20}$,
M.~Baszczyk$^{\,14}$,
S.~Bhosale$^{\,11}$,
A.~Blondel$^{\,4}$
% ...

\end{sloppypar}
\bigskip
% based on XML DB with time Wed Jul 14 09:19:03 2021
% Institutes in alphabetical order.
\scriptsize
\noindent
% ...
$^{3}$~Ru{\dj}er Bo\v{s}kovi\'c Institute, Zagreb, Croatia\\
$^{4}$~LPNHE, University of Paris VI and VII, Paris, France\\
$^{5}$~Karlsruhe Institute of Technology, Karlsruhe, Germany\\
% ...
$^{10}$~Jan Kochanowski University in Kielce, Poland\\
$^{11}$~Institute of Nuclear Physics, Polish Academy of Sciences, Cracow, Poland\\
% ...
$^{14}$~AGH -- University of Science and Technology, Cracow, Poland\\
% ...
$^{20}$~Joint Institute for Nuclear Research, Dubna, Russia\\
% ...
$^{22}$~St.~Petersburg State University, St.~Petersburg, Russia\\
% ...
$^{26}$~University of Colorado, Boulder, USA\\
% ...

\end{document}